# β-Ga$_2$O$_3$ Trench Schottky Diodes by Novel Low-Damage Ga-Flux Etching

Sushovan Dhara[a], Nidhin Kurian Kalarickal[a], Ashok Dheenan[a], Sheikh Ifatur Rahman[a], Chandan Joishi[a], Siddharth Rajan[a,b]

[a] *Department of Electrical and Computer Engineering, The Ohio State University, USA,*
[b] *Department of Materials Science and Engineering, The Ohio State University, USA.*
*Email: dhara.2@osu.edu, rajan@ece.osu.edu*

**Abstract:** β-Ga$_2$O$_3$ trench Schottky barrier diodes fabricated through a Gallium atomic beam etching technique, with excellent field strength and power device figure of merit, are demonstrated. Trench formation was accomplished by a low-damage Ga flux etch that enables near-ideal forward operating characteristics that are independent of fin orientation. The reverse breakdown field strength of greater than 5.10 MV/cm is demonstrated at breakdown voltage as of 1.45 kV. This result demonstrates the potential for Ga atomic beam etching and high-quality dielectric layers for improved performance in β-Ga$_2$O$_3$ vertical power devices.

β-Ga$_2$O$_3$ is a promising material for kV class power devices due to its high theoretical breakdown field (8 MV/cm) and power switching figure of merit[1–3]. The availability of melt-based native substrates can provide semiconductor devices with lower defects and greater reliability at a lower cost[4,5]. Unlike other vertical power devices based on SiC and GaN, the unavailability of p-type β-Ga$_2$O$_3$[6] makes power device design challenging. Without the ability to form p-n homojunctions, diode architectures can only be realized using Schottky junctions. However, this also poses a limitation in device breakdown performance, since the Schottky barrier used for rectification cannot sustain high fields in reverse bias. The narrow triangular tunnelling barrier near the metal-semiconductor interface under high field results in high tunnelling current into the semiconductor and premature breakdown. This Schottky junction limits the material's potential to be used in high-voltage diode applications. The use of alloyed interlayers such as AlGaO[7], which possesses a higher Schottky barrier, or the use of high-k dielectric such as BaTiO$_3$[8] heterojunction to maintain a wider tunneling barrier showed significant improvement to in sustaining high fields. Another method to design efficient diodes is through formation of three-dimensional 'trench' device geometries, where the peak electric field at the metal-semiconductor interface is pushed away into the bulk of the semiconductor.

This approach can provide an excellent reverse breakdown field without sacrificing a forward operating voltage close to the Schottky turn-on[9,10].

To realize the trench Schottky barrier diode, it is critical to reduce damage to the semiconductor in the fin etching process module. Additionally, an appropriate dielectric with a high breakdown strength needs to be integrated. It is well-known that ICP/RIE etching causes sub-surface damage and depletion in $\beta$-$Ga_2O_3$[11,12]. This report demonstrates a novel approach to realize high aspect ratio trenches through a combination of low-damage Ga atomic flux-based etching process and an $Al_2O_3$ dielectric layer deposited by plasma-assisted MBE to fabricate trench Schottky diodes. Exposing $\beta$-$Ga_2O_3$ to Ga flux in vacuum results in a purely chemical, damage-free etching process that significantly improves the on-resistance when compared to devices with trenches formed by ICP/RIE processes.

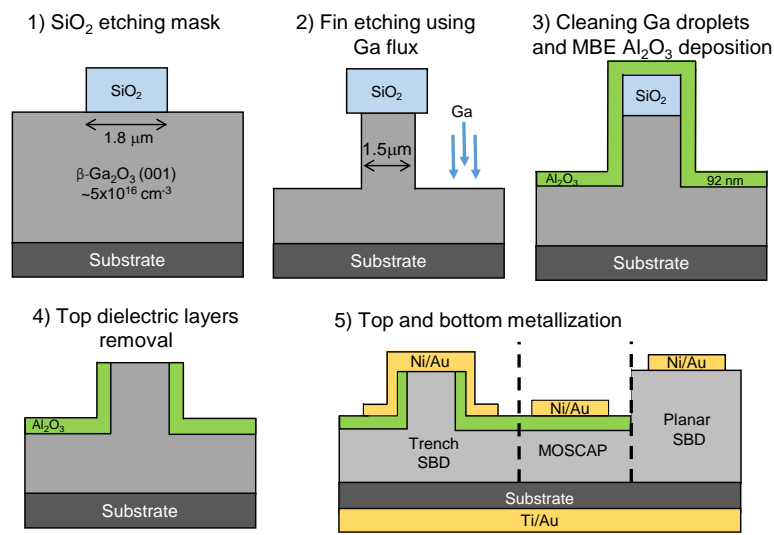

*Figure 1: Process flow for the Ga-flux etched trench SBD*

The devices discussed here were fabricated on commercially obtained HVPE-grown $\beta$-$Ga_2O_3$ epitaxial layers (10 $\mu m$ thick epitaxial layer of doping ($N_D$) ~ $5\times10^{16}$ $cm^{-3}$) sourced from Novel Crystal Technologies, Japan. The process flow (figure 1) began with the definition of the 100 μm x 1.8 μm stripes to define the trench geometries. 100 nm of $SiO_2$ was blanket deposited for use as a hard-mask using plasma-enhanced chemical vapor deposition (PECVD) at 250°C and then patterned for fin etching using photolithography, followed by dry and wet etching of SiO$_2$. The β-Ga$_2$O$_3$ fins were then defined by exposing the sample to atomic gallium flux[13], where reaction of Ga with the β-Ga$_2$O$_3$ leads to the formation of volatile sub-oxide (Ga$_2$O) which desorbs at temperatures above 500°C, resulting in chemical etching of the sample. A two-step etching process (step 1- T$_{sub}$=700 °C, BEP = 5x10$^{-7}$ Torr, time = 120 minutes and step 2- T$_{sub}$=550 °C, BEP = 1.5x10$^{-7}$ Torr, time = 35 minutes) was carried out in a Riber M7 molecular beam epitaxy (MBE) chamber. This two-step process helps to avoid the formation of a Si-rich layer near the top of the etched material[13]. The process results in sharply defined features with smooth sidewalls. The trench depth was measured by atomic force microscopy to be ~1.2 µm. After the Ga-flux etching process, the sample was taken out of the MBE chamber and dipped in HCl for 5 minutes to remove any excess Ga droplets from the surface. After the etching step, the sample was reintroduced into the MBE chamber, a a 92 nm thick Al$_2$O$_3$ layer was deposited over the entire sample by exposing the sample to atomic Al-flux and oxygen plasma (Al flux = 2x10$^{-8}$ Torr, O$_2$ flow=2.5 sccm, plasma power=250 W) at a substrate temperature of 400°C in the same MBE chamber. Windows to the top of each fin were then patterned by photolithography and the SiO$_2$ and Al$_2$O$_3$ dielectric layers were removed by a dry etching, followed by wet etching (dilute buffered oxide etch (BOE)). Top Schottky contacts were then formed by sputtering 50 nm of Ni for side wall coverage followed by electron beam evaporation of Ni/Au (30/100 nm). Finally, the ohmic back contact (Ti/Au, 30/70 nm) was deposited by electron-beam evaporation. MOS-capacitors were fabricated on the sample sample on the Ga-flux etched β-Ga$_2$O$_3$ surface. Additionally, a planar SBD on the un-etched β-Ga$_2$O$_3$ was fabricated for comparison.

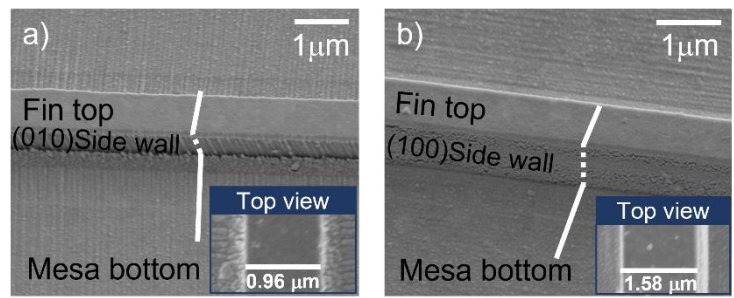

*Figure 2 : SEM imaging of (a) [100]-direction and, (b) [010]-direction-oriented fin structure*

Fin structures were fabricated along both the [100] and [010] directions to investigate the effect of Ga flux-based etching on different crystal planes. Figure 2 shows the tilted SEM of the fin structures and the top view of the fins oriented along the [010] and [100] directions. Both the fins were etched using a 1.8 μm wide $SiO_2$ mask, and 1.2 μm of vertical etching resulted in sub-lithographic fin widths of 1.58 μm and 0.96 μm along [010] and [100] directions, respectively.

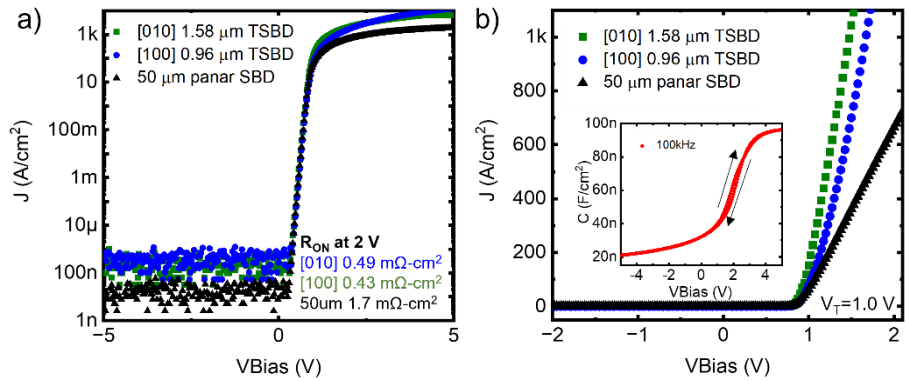

*Figure 3: Forward current-voltage characteristics of the planar and trench SBDs (TSBD) in (a) log and (b) linear scale with capacitance-voltage characteristics of the MOS capacitor on the etched surface in the inset.*

Two terminal electrical characterization was carried out using a Keysight B1500A semiconductor parameter analyzer. Room temperature current-voltage (J-V) measurements of the two types of fin device structures and large-area planar Schottky diodes are shown in figure 3(a-b). The forward J-V characteristics of both the trench Schottky barrier diodes showed a turn-on voltage of ~1.0 V, similar to the turn-on voltage of a planar Schottky diode fabricated on the same epitaxial structure. The trench SBDs oriented along [100] and [010] directions showed similar turn-on voltage and and resistance. In comparison with fins reported previously that were fabricated using dry etching exhibited depletion of carriers in the drift layer from plasma damage and strong dependence on fin orientation[11,12]. The specific on-resistance of 50 μm circular planar device was ~1.7 mΩ-cm², while trench diodes showed on-resistance close to 0.5 mΩ-cm² at 2 V of forward biasing. This lower resistance of fin structures is due to current spreading. When we account for the current spreading, the on-resistance is computed to be 1.01 mΩ-cm² and 1.20 mΩ-cm² for fins oriented in [100] and [010] direction (details in supplementary section). The reverse leakage current was found to be close to the noise floor margin of the measurement system for both the planar and trench SBDs. To understand the properties of the $Al_2O_3$ dielectric layers and the interface with the semiconductor, MOS capacitor structures on the same sample were electrically characterized. The capacitance-voltage (CV) characteristics of the MOS-capacitor showed relatively low voltage hysteresis (figure 3(b)-inset), indicating a good quality interface. When compared with the simulated ideal CV of this MOS structure, a positive flat band voltage shift was observed. This confirms the presence of negative fixed charge either at the $Al_2O_3/\beta$-$Ga_2O_3$ interface or in the bulk of the $Al_2O_3$.

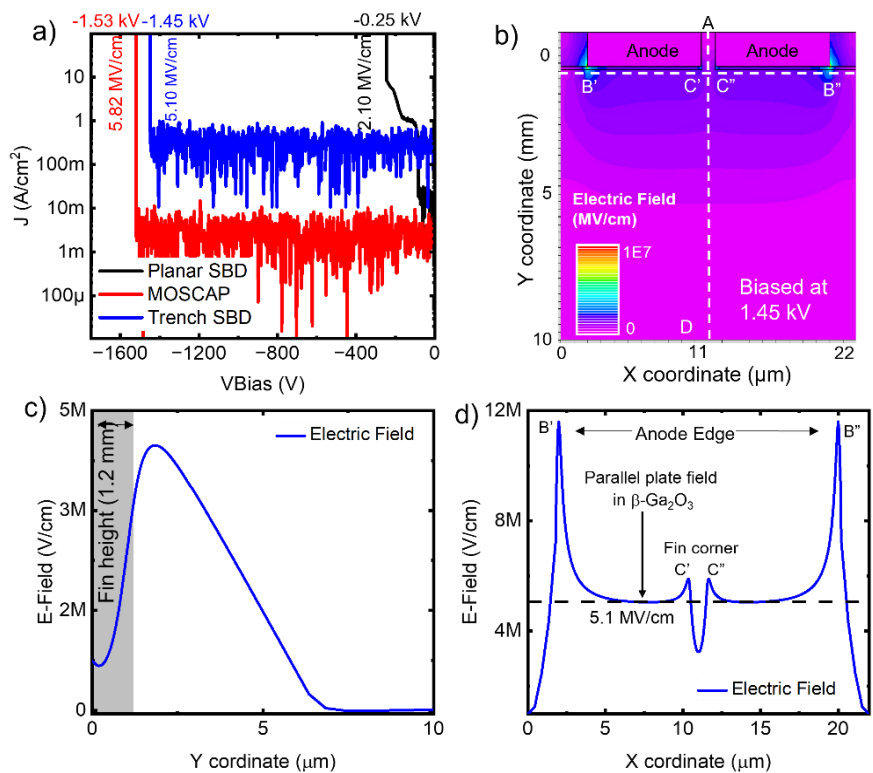

*Figure 4: (a) Reverse breakdown measurement of the MOSCAP, planar, and trench SBDs. (b) Simulated 2-dimensional electric field profiles of 0.98 μm wide fin at 1.45 kV. Electric field profile along (c) AD (vertical cutline) and (d) B'C'C"B" (lateral cutline) for 0.96 μm wide fin.*

The devices whose forward characteristics had been measured were selected for measurement of higher reverse voltage breakdown using a Agilent B1505A semiconductor parameter analyzer. The reverse J-V measurements show hard (destructive) breakdown characteristics for the highest obtained reverse breakdown votlage from planar Schottky diodes

(250 V), trench SBD (1.45 kV), and a planar MOS capacitor (1.52 kV) (figure 4(a)). To investigate the electrostatics of the trench SBD, 2D simulations of the device structures were done using the Silvaco ATLAS package. Results for the 0.96 μm-width fin are shown for an applied voltage of 1.45 kV in figure 4(b). The vertical cutline A-D along the structure shows the electric field profile along the center of the fin (figure 4(c)). The electric field at the metal/$\beta$-Ga$_2$O$_3$ Schottky interface is <2 MV/cm. Along the vertical cutline, the peak electric field was observed in the bulk at a depth of ~2 μm, as expected from the trench architecture. The lateral cutline as shown in figure 4(d) along B'-C'-C"-B" shows the presence of a peak in the electric field near the anode edge at points B' and B". This suggests that peak fields at the metal edge (approaching > 8 MV/cm) are likely the cause of the breakdown.

We estimated the parallel plate fields in the $\beta$-Ga$_2$O$_3$ at the reverse breakdown condition to be 2.10 MV/cm, 5.10 MV/cm, and 5.82 MV/cm for the planar SBD, trench SBDs, and the MOSCAP respectively. For planar structures, the breakdown field was estimated using one-dimensional electrostatics, $E_{field} = \sqrt{qN_D V_{BR}/\varepsilon}$, where $q$ is charge of electron, $N_D$ is the doping of the semiconductor, $V_{BR}$ is the breakdown voltage, and $\varepsilon$ is the permitivity of the semiconductor. For the trench structure, we estimated the field from simulation, as shown in Figure 4(d). Observation of the devices post-breakdown suggest catastrophic breakdown near the anode edge, which matches the peak field position expected from simulations.

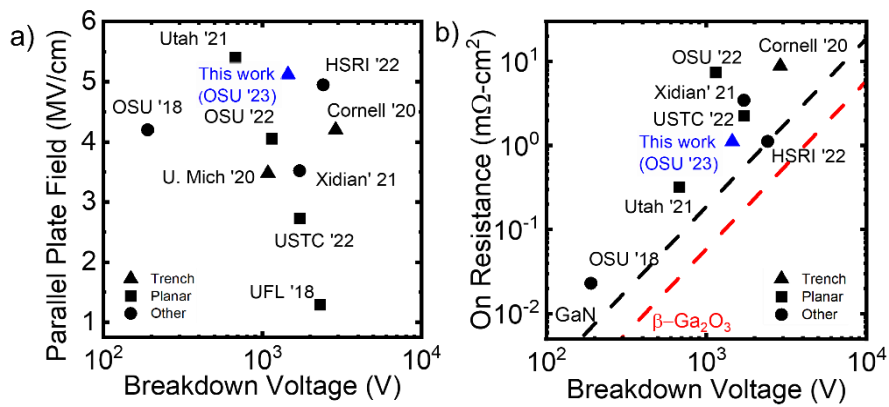

*Figure 5. (a) Parallel-plate field vs breakdown voltage voltage of Ga-flux etched trench SBD in comparison to previously reported devices. (b) Plot of on-resistance vs. breakdown*

The breakdown voltage and parallel plate electric field of the trench Schottky diode reported here is compared with previously reported $β$-Ga$_2$O$_3$ vertical diodes in Figure 5. [10,14–20]. The device demonstrated here shows one of the highest combinations of breakdown electric field strength and breakdown voltage. Furthermore, the damage-free atomic beam etching method adopted here leads to excellent forward conduction characteristics when compared to standard ICP/RIE processes. The evaluated Baliga power device figure of merit (BFOM) is also benchmarked with other reports in figure 5(b) and estimated to be higher than 2 GW/cm$^2$, which is one of the best values reported for $β$-Ga$_2$O$_3$, and compares favorably with state-of-art SiC vertical devices. Improved edge termination is expected to greatly improve the performance of the devices.

In summary, we have demonstrated a novel process to realize $β$-Ga$_2$O$_3$ trench Schottky diodes using Ga atomic beam etching technique. A parallel-plate electric field strength of 5.12 MV/cm with an associated breakdown voltage of ~1.45 kV combined with a low on-resistance results in a BFOM of greater than 2 GW/cm$^2$. This result shows the potential of $β$-Ga$_2$O$_3$ vertical device structures for future high voltage electronics. We acknowledge funding from Department of Energy / National Nuclear Security Administration under Award Number(s) DE-NA0003921, and AFOSR GAME MURI (Award No. FA9550-18-1-0479, project manager Dr. Ali Sayir). The content of the information does not necessarily reflect the position or the policy of the federal government, and no official endorsement should be inferred.

# Supporting Information

# β-Ga$_2$O$_3$ Trench Schottky Diodes by Novel Low-Damage Ga-Flux Etching

Sushovan Dhara[a], Nidhin Kurian Kalarickal[a], Ashok Dheenan[a], Sheikh Ifatur Rahman[a], Chandan Joishi[a], Siddharth Rajan[a,b]

[a] Department of Electrical and Computer Engineering, The Ohio State University, USA,
[b] Department of Materials Science and Engineering, The Ohio State University, USA.
Email: dhara.2@osu.edu, rajan@ece.osu.edu

Supporting Information 1: (Estimation of On-resistance)

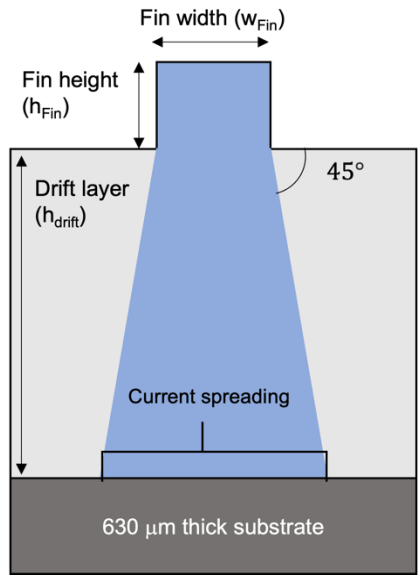

Device schematic for On-resistance estimation.

$$Resistivity\ (\rho) = \frac{1}{q\mu N_D}.$$

Where, $N_D$ is Doping of the semiconductor, q is electron charge, and μ is mobility

Thus,

$$\rho_{epi.} = 1.25\ \Omega cm\ \left(N_D = \frac{5E16}{cm^3},\ \mu = \frac{100\ cm^2}{Vs}\right)$$

and,

$$\rho_{substrate} = 10.8\ m\Omega cm\ \left(ND = \frac{5.8E18}{cm^3},\ \mu = \frac{100\ cm^2}{Vs}\right)$$

$R_{substrate} = \rho_{substrate} \times substrate\ thicknes = 0.68\ m\Omega cm^2$

$R_{epi.} = \rho_{epi.} \times w_{Fin} \times \left(\left(\frac{h_{Fin}}{w_{Fin}}\right) + \frac{1}{2}\ln\left(1 + \frac{2h_{drift}}{w_{Fin}}\right)\right)$, by assuming current spreading angle of 45°

The total device resistance can be estimated as,

$$R_{Device} = R_{substrate} + R_{epi.}$$

$R_{epi.[100]}= 0.33\ m\Omega cm2 \rightarrow R_{Device[100]}= 1.01\ m\Omega cm^2$

$R_{epi.[010]}= 0.52\ m\Omega cm2 \rightarrow R_{Device[010]}= 1.20\ m\Omega cm^2$

$R_{epi(50\ \mu m)}= 1.05\ m\Omega cm2 \rightarrow R_{Device(50\ \mu m)}= 1.73\ m\Omega cm^2$

This estimated resistance of the 50 μm device is close to the measured resistance. Thus, we can consider this resistance estimation assumption for trench devices.